\input{aipcheck}

\documentclass[
    ,final            
]
  {aipproc}
\usepackage{subfigure}
\layoutstyle{6x9}

\begin{document}

\title{Azimuthal di-hadron correlations in Au+Au collisions at
$\sqrt{s_{NN}}=200$ GeV from STAR}

\classification{25.75 Dw}\keywords{heavy-ion collisions, jet quenching, di-hadron correlations}
\author{Mark J. Horner (for the STAR Collaboration\footnote{http://www.star.bnl.gov/central/collaboration/authors/authorList.php} )}{
  address={Nuclear Science Division, MS 70R319, Lawrence Berkeley National Lab, Berkeley, CA 94720}
}

\begin{abstract}
We present the centrality dependence of the azimuthal angle correlation shapes between a trigger and associated 
charged hadrons in Au+Au collisions at $\sqrt{s_{NN}}=200$ GeV from the STAR experiment. The away-side structure
is significantly modified in central collisions and is characterised in terms of a doubly-peaked yield. A minimum in the
<$p_T$> of hadrons opposite to the trigger particle is also investigated.
\end{abstract}

\maketitle

Ultra-relativistic heavy-ion collisions were expected to create 
a weakly-interacting, deconfined state of quarks and gluons.
Results from RHIC 
have shown that bulk particle production can be well modelled using hydrodynamics and the system may be strongly
coupled. Di-hadron correlation studies~\cite{Disappearance,Resurrection} have shown
that hard partons from initial-state, back-to-back scatterings interact strongly
with the 
system that is created, and can be used as a probe of the medium.

We present further di-hadron correlation results which are consistent
with a picture in which hard-scattered partons, propagating faster
than the speed of sound, deposit energy in the medium to set
up conical flow.

We investigated di-hadron number correlations by choosing a trigger charged
hadron in a particular $p_T$ range and constructing a distribution of azimuthal angular
differences between the trigger and associated charged hadrons in a specific $p_T$ range. All
hadrons were restricted to the range -1.0<$\eta$<1.0.
The associated 
$p_T$ window may overlap with that of the trigger hadrons. All presented results are corrected for
the single particle efficiency and acceptance as well as the pair acceptance as a function of azimuthal difference.

The measured correlations contain a background
of uncorrelated particles. 
The correlation of all particles with the reaction plane through elliptic flow, $v_2$~\cite{v2PRC},
gives rise to a $\cos{2\Delta\phi}$ modulation of the background distribution.
The background is removed by 
normalising the level of the
$v_2$-modulated background to the minimum value of the raw correlation. 
The nominal $v_2$ value we
subtract is the mean of two measurements~\cite{v2PRC,v2_4part} using the reaction
plane and four-particle cumulant techniques which give
different results for $v_2$ due to different sensitivities to non-flow effects. The difference between the two limiting $v_2$ results is used as the
estimate of the systematic uncertainty. 

\begin{figure}[!htb]
\includegraphics[scale=0.35]{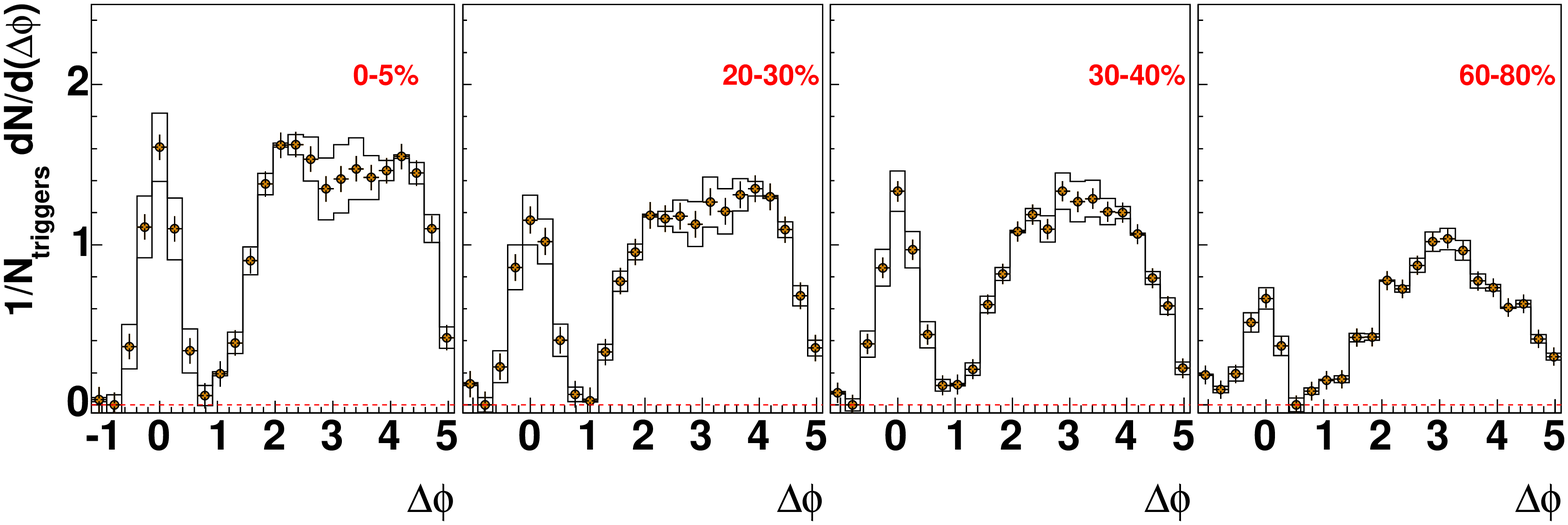}
  \caption{Centrality dependence of number correlations for trigger hadrons with $p_T$ between 4.0 and 6.0 GeV/c associated
  with hadrons with a $p_T$ between 0.150 and 4.0 GeV/c. The panels show from
  top to bottom, left to right, decreasing centrality. Black lines show systematic uncertainty from the $v_2$ modulation of the background. 
  The horizontal axis is offset for clarity.}
  \label{fig:cent}
\end{figure}

Figure~\ref{fig:cent} shows the background-subtracted azimuthal number
correlations for various centralities for trigger particles with $p_T$ between
4.0 and 6.0 GeV/c and associated particles with $p_T$ between 0.15 and 4.0 GeV/c. 
While for the most peripheral events (60-80\%) two clear peaks are visible at $\Delta\phi=$0 and $\pi$, the 
away-side structure is flatter and broader for more central events. The near-side yield also
increases with increasing centrality and this may be related to a ridge-like yield in $\Delta\eta$~\cite{Joern}.
This behaviour
is also observed in other trigger $p_T$ ranges. 

To characterise the away-side shape, we have fitted the distribution with several functional forms based 
on the idea that there is significant yield in a cone around $\pi$. This part of the yield is parameterised 
by a pair of Gaussian distributions symmetric about $\pi$. In addition some jet yield may still exist 
which is centered at $\pi$ and we have characterised this using a $\cos{\Delta\phi}$ or Gaussian distribution.

\begin{figure}[!htb]
  \includegraphics[scale=0.4]{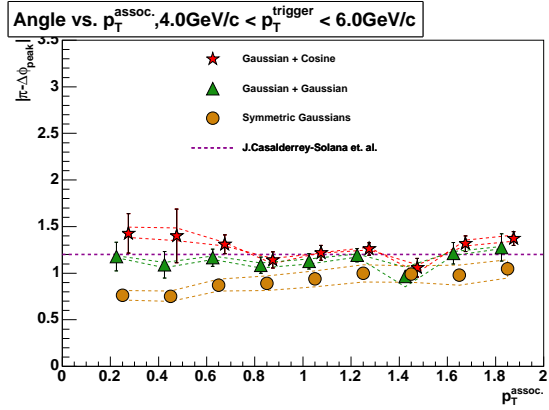}
  \caption{Distance of the 
  symmetric Gaussian away from $\pi$ as a function of associated particle $p_T$ for trigger particles with $p_T$s between 4.0 and 6.0 GeV/c
  in 0-5\% central collisions. Horizontal line inspired by \cite{sh}. The dashed bands show the systematic uncertainty from $v_2$ determination.}
  \label{fig:location}
\end{figure}

Figure~\ref{fig:location} shows half the separation between the Gaussian yields not centered at $\pi$ as a
function of associated hadron $p_T$. 
It can be seen in the figure that the separation is independent of the choice of central distribution shape and approximately
independent of associated $p_T$, while if no additional yield is assumed to be present at $\Delta\phi=\pi$ (circle markers) the separation is slightly
smaller and increasing with $p_T$.
These results are not consistent with a picture
in which the fragmentation of Cherenkov gluons is the dominant production mechanism for the yield away from $\Delta\phi=\pi$~\cite{abhijit}.

\begin{figure}[!htb]
  \hfill
  \begin{minipage}[t]{.5\textwidth}
    \begin{center}
  \includegraphics[scale=0.38]{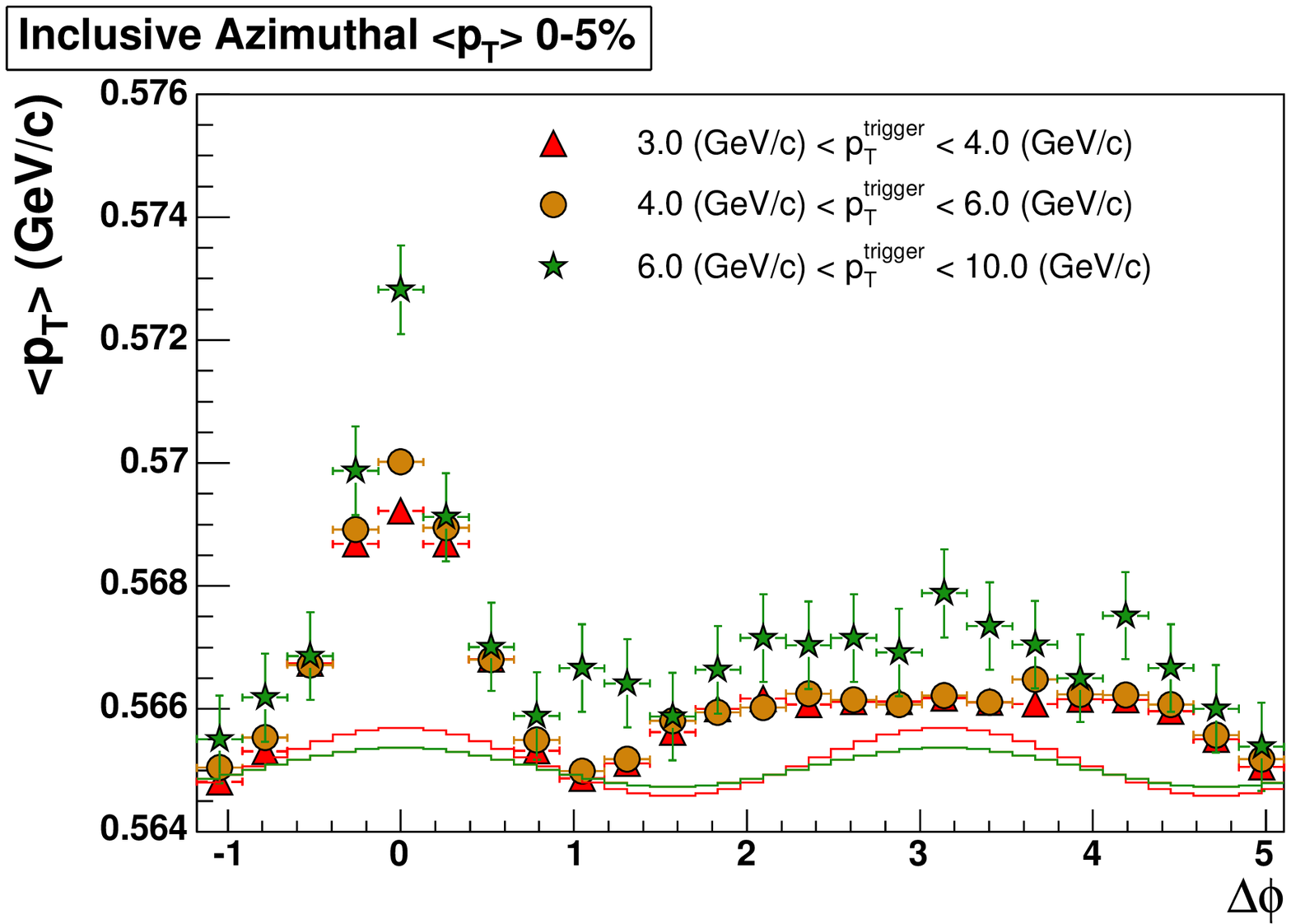}
    \end{center}
  \end{minipage}
  \hfill
  \begin{minipage}[t]{.5\textwidth}
    \begin{center}
  \includegraphics[scale=0.38]{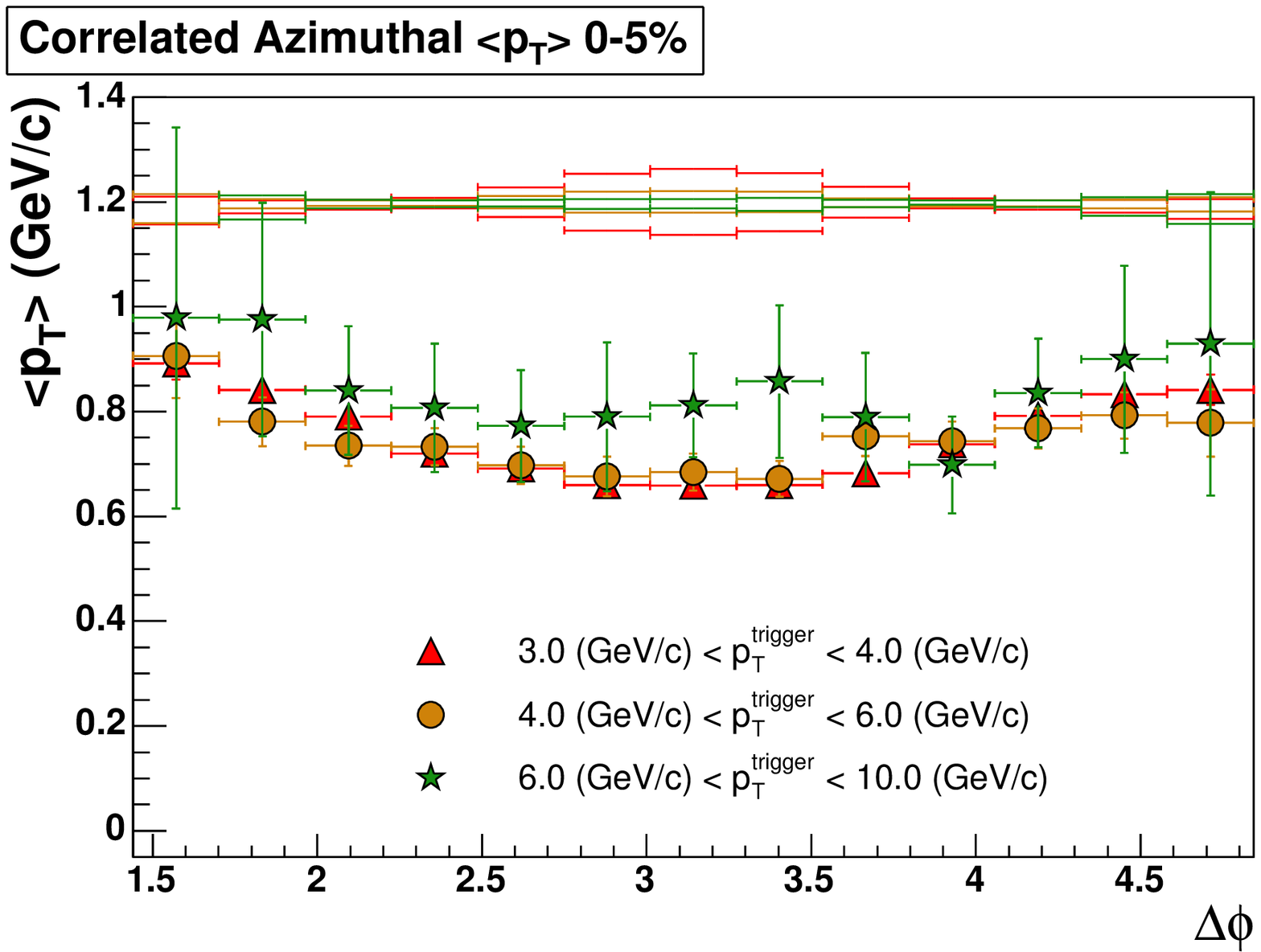}
    \end{center}
  \end{minipage}
  \caption{
  Inclusive, azimuthal <$p_T$> (left panel) of associated hadrons in all events containing
  various classes of trigger particles. The lines show the variation in background for the different
  trigger ranges. In the right panel the away-side for the azimuthal <$p_T$> of
  correlated particles. The <$p_T$> is calculated for associated particles between
  0.150 and 4.0 GeV/c. The lines show the systematic uncertainty in $v_2$.}
  \hfill
  \label{fig:meanpt}
\end{figure}

Furthermore, we investigated
the azimuthal dependence of associated hadron mean $p_T$, <$p_T$>.
Figure~\ref{fig:meanpt} shows the <$p_T$> as a function of azimuth for three different ranges of trigger $p_T$.
The near-side <$p_T$> values exhibit the expected hierarchy with increasing trigger $p_T$.
The away-side inclusive
<$p_T$> distributions are similar for the two softer trigger ranges and slightly increased
for the hardest triggers. For the
correlated (background-subtracted), associated hadron <$p_T$> we focus on the away-side in the right panel of Fig.~\ref{fig:meanpt}. The results show a
distribution depleted at $\pi$ for the two softer trigger cases.
For the case of trigger particles with $p_T$ in the range 6.0 to 10.0 GeV/c the results
are consistent with both depleted and flat distributions within
statistical errors.

In summary, a significant broadening consistent with the appearance of secondary peaks 
has been observed in central Au+Au collisions
as $\sqrt{s_{NN}}=200$ GeV. The interpretation of this is still under investigation but the observed behaviour is
consistent with conical flow built up in the medium. The angle of these secondary peaks relative to the center of the 
away-side distribution is approximately
independent of associated hadron $p_T$. Similar structures are seen in associated hadron <$p_T$> as a function of azimuthal angle
with respect to the trigger particle.


\end{document}